\documentclass[a4paper,11pt]{article}
\usepackage{pos}
\usepackage{physics}
\usepackage{dsfont}
\usepackage{bbm}
\usepackage{hyperref}

\title{Using classical bit-flip correction for error mitigation in quantum computations
including 2-qubit correlations}
\ShortTitle{Using classical bit-flip correction for error mitigation including 2-qubit correlations}

\author[a,b]{Constantia Alexandrou}
\author[c,d]{Lena Funcke}
\author[e]{Tobias Hartung}
\author*[f]{Karl Jansen}
\author[b]{Stefan K\"uhn}
\author[a,b]{Georgios Polykratis}
\author[f,g]{Paolo Stornati}
\author[h]{Xiaoyang Wang}

\affiliation[a]{Department of Physics, University of Cyprus, P.O. Box 20537, 1678 Nicosia, Cyprus}

\affiliation[b]{Computation-based  Science  and  Technology  Research  Center, The  Cyprus  Institute,  20  Kavafi  Street,  2121  Nicosia, Cyprus}

\affiliation[c]{Center for Theoretical Physics, Co-Design Center for Quantum Advantage, and NSF AI Institute for Artificial Intelligence and Fundamental Interactions, Massachusetts Institute of Technology, 
77 Massachusetts Avenue, Cambridge, MA 02139, USA}

\affiliation[d]{Perimeter Institute for Theoretical Physics, 31 Caroline Street North, Waterloo, ON N2L 2Y5, Canada}

\affiliation[e]{Department of Mathematical Sciences, University of Bath, 4 West, Claverton Down, Bath, BA2 7AY, UK}

\affiliation[f]{Deutsches Elektronen-Synchrotron DESY,  Platanenallee 6, 15738 Zeuthen}

\affiliation[g]{Institut für Physik, Humboldt-Universität zu Berlin, Zum Großen Windkanal 6, D-12489 Berlin, Germany}

\affiliation[h]{School of Physics, Peking University, 5 Yiheyuan Rd, Haidian District, Beijing 100871, China}

\emailAdd{alexand@ucy.ac.cy}
\emailAdd{lfuncke@mit.edu}
\emailAdd{tobias.hartung@desy.de}
\emailAdd{karl.jansen@desy.de}
\emailAdd{s.kuehn@cyi.ac.cy}
\emailAdd{g.polykratis@cyi.ac.cy}
\emailAdd{paolo.stornati@desy.de}
\emailAdd{zzwxy@pku.edu.cn}

\abstract{We present an error mitigation scheme which corrects 
readout errors on Noisy Intermediate-Scale Quantum (NISQ) computers~\cite{Funcke:2020olv,Funcke:2021aps}. After a short review of applying the 
method to one qubit, we proceed to discuss the case 
when correlations between different qubits occur. We demonstrate how the readout error can be mitigated
in this case. By performing 
experiments on IBMQ hardware, we show that such correlations do not 
have a strong effect on the results, justifying to neglect them.\\

Preprint number: MIT-CTP/5352} 

\FullConference{%
 The 38th International Symposium on Lattice Field Theory, LATTICE2021
  26th-30th July, 2021
  Zoom/Gather@Massachusetts Institute of Technology
}

\begin{document}
\maketitle

\section{Introduction}

In recent years, 
quantum computing has started to become a very active area of research 
in lattice field theory. The reason  
is that quantum computations offer the exciting possibility 
to solve problems which are either extremely hard or even 
impossible to address on classical computers. This includes 
systems with a non-zero chemical potential, topological terms, and real-time 
evolutions. 

This promising avenue is, however, blocked by the fact that current 
quantum computers, so-called Noisy Intermediate-Scale Quantum (NISQ) computers, 
only have a small number of qubits which, in addition, are very noisy. This leads to various kind of errors in quantum computations, 
which often prevent obtaining solutions with a desired 
accuracy. 
However, already in current and near-term NISQ devices, these errors can be partly corrected through 
error mitigation schemes (see, e.g., Refs.~\cite{Cramer_2016,Kandala:2017,Li:2017,Temme:2017,Endo:2018,Endo:2019,Kandala:2019,Tannu:2019,YeterAydeniz2019,YeterAydeniz2020,Funcke:2020olv,Funcke:2021aps,chen2021exponential,berg2021modelfree,Geller2021}).
These works include various ideas to mitigate the noise, e.g., 
by altering the quantum circuit, by post-processing the data, 
or by measuring  modified operators. 
As demonstrated in the above references, in this way the quantum noise can indeed be 
mitigated leading to more reliable estimates of physical results. 
Another important improvement 
is the construction of minimal but maximally expressive quantum 
circuits, which has been developed by some us 
in Refs.~\cite{Funcke_2021,funcke2021bestapproximation}.

\section{Readout error mitigation} 

In Refs.~\cite{Funcke:2020olv,Funcke:2021aps}, we have developed a general error 
mitigation scheme for readout errors. Our method is based on a readout 
error calibration of qubits and a reinterpretation of the measurements as 
measuring ``noisy operators''. In particular, we have demonstrated that our 
error mitigation method scales efficiently, i.e., polynomially. 
Our method can be applied either as a pre-processing or as a post-processing step. 
Further experiments and a comparison of our error mitigation scheme
on IBMQ and Rigetti hardware is provided in Ref.~\cite{Alexandrou:2021ynh} in this 
conference. 
As we will show below and discussed in Ref.~\cite{Funcke:2020olv},  
the method can also take correlations between qubits into account. We will 
demonstrate this both theoretically and in practical experiments. 
In order to explain the main idea of the method, we will start, however, with 
a description where correlations are neglected. 

\subsection{Neglecting correlations} 

Let $p_{q, 1}$ be the probability of erroneously reading out a qubit $q$ such that an incorrect outcome~$1$ is measured instead of a correct outcome~$0$, and $p_{q, 0}$ the probability of erroneously reading out $0$ instead of $1$. These probabilities can be obtained by preparing each qubit in the computational basis states and measuring the outcomes (see Fig.~\ref{fig:calibration_circuits}). For example, to obtain $p_{q,1}$ we prepare the qubit~$q$ multiple times in the state $\ket{0}$ and record the number of outcomes $1$.
Then, if we want to measure an operator, e.g., the $Z_2\otimes Z_1$ operator on 2 qubits (where $Z$ represents 
the third Pauli matrix $\sigma_Z$), the correct expectation value for $Z_2\otimes Z_1$ 
can be determined 
by measuring the operator $Z_2\otimes Z_1$ itself, the operator $Z_2\otimes \mathbbm{1}_1$, and the 
operator $\mathbbm{1}_2\otimes Z_1$. From the measurement of the probabilities $p_{q,0/1}$
in the calibration process, we can obtain the coefficients 
of these operators, and we find 
\begin{align}
    \begin{aligned}
        Z_{2} \otimes Z_{1}=\,&
        \frac{1}{\gamma\left(Z_{2}\right) \gamma\left(Z_{1}\right)} \mathbb{E}\left(\tilde{Z}_{2} \otimes \tilde{Z}_{1}\right)
        -\frac{\gamma\left(\mathbbm{1}_{1}\right)}{\gamma\left(Z_{2}\right) \gamma\left(Z_{1}\right)} \mathbb{E}\left(\tilde{Z}_{2}\right) \otimes \mathbbm{1}_{1} \\ &
        -\frac{\gamma\left(\mathbbm{1}_{2}\right)}{\gamma\left(Z_{2}\right) \gamma\left(Z_{1}\right)} \mathbbm{1}_{2} \otimes \mathbb{E}\left(\tilde{Z}_{1}\right)
        +\frac{\gamma\left(\mathbbm{1}_{2}\right) \gamma\left(\mathbbm{1}_{1}\right)}{\gamma\left(Z_{2}\right) \gamma\left(Z_{1}\right)} \mathbbm{1}_{2} \otimes \mathbbm{1}_{1}\, ,
    \end{aligned}
    \label{eq:inverted}
\end{align}
\noindent where the tilde denotes a noisy operator measured on noisy quantum hardware, $\mathbb{E}$ denotes the expectation value of the noisy operator subject to bit flips, 
which should not be confused with the quantum mechanical expectation value, and we defined
\begin{align}
    \gamma\left(O_{q}\right)&:= \left\{\begin{array}{ll}1-p_{q, 0}-p_{q, 1} & \text { for } O_{q}=Z_{q} \\
    p_{q, 1}-p_{q, 0} & \text { for } O_{q}=\mathbbm{1}_{q} .\end{array}\right\}\; .
    \label{eq:gammas}
\end{align}
Thus, 
the correct expectation value of a two-qubit operator can be obtained 
by measuring noise-afflicted expectation values of $Z_1$, $Z_2$, and $Z_1\otimes Z_2$ 
on the quantum device and by combining them with coefficients that only 
depend on the known bit-flip probabilities.
A key element is that $\mathbb{E}\left(\tilde{Z}_{Q} \cdots \tilde{Z}_{1}\right)$ can be 
factorized into single-qubit expectation values
$\mathbb{E}\left(\tilde{Z}_{Q} \cdots \tilde{Z}_{1}\right)=\mathbb{E} \tilde{Z}_{Q} \cdots \mathbb{E} \tilde{Z}_{1}\, .
$
The iterative proof of this equation is given in Ref.~\cite{Funcke:2020olv} and eventually leads 
 to the fact that the readout error mitigation method scales polynomially. 

\subsection{Taking correlations into account} 

\subsubsection{Theoretical framework}

The construction of the error-mitigated $Z_2\otimes Z_1$ operator for 2 qubits in Eq.~\eqref{eq:inverted} is based on a probabilistic description at the operator level. For example, if a single $Z_q$ operator is measured on qubit $q$, then a bit-flip of $1\to0$ and $0\to0$  occurs with a probability of $p_{q,1}(1-p_{q,0})$. In this case, the measurement outcome is always $0$, such that measuring $Z_q$ is equivalent to measuring the identity operator $\mathbbm{1}$ with probability $p_{q,1}(1-p_{q,0})$. Considering all four bit-flip cases yields a probability distribution of random operators $\tilde Z_q$. The expectation of $\tilde Z_q$ with respect to this probability distribution is 
\begin{align}\label{eq:1Q-EZ}
  \mathbb{E}(\tilde Z_q)=(1-p_{q,0}-p_{q,1})Z_q+(p_{q,1}-p_{q,0})\mathbbm{1}_q.
\end{align}
Rearranging Eq.~\eqref{eq:1Q-EZ} yields an expression for the operator $Z_q$ that we wish to measure, in terms of constants depending on $p_{q,0/1}$ and of the expectation of $\tilde Z_q$ subject to the bit-flip distribution measured on the quantum device. If no correlations between qubits exist, then we can build the two-qubit result of Eq.~\eqref{eq:inverted} by tensoring the expressions for $Z_q$ obtained from Eq.~\eqref{eq:1Q-EZ} on both qubits.

In the presence of inter-qubit correlations, a similar method can be employed. For example, if we wish to measure the operator $Z_2\otimes Z_1$, then we can calibrate the probabilities $p(b|b')$ of finding the bitstring $b$ given that the underlying bitstring was $b'$, i.e., $p(10|00)$ is the probability of flipping the second qubit from $0$ to $1$ and keeping the first qubit in $0$. We then compute the expected operators $\mathbb{E}(\widetilde{Z_2\otimes Z_1})$, $\mathbb{E}(\widetilde{Z_2\otimes\mathbbm{1}_1})$, and $\mathbb{E}(\widetilde{\mathbbm{1}_2\otimes Z_1})$ by considering the induced probability distribution, e.g.,
\begin{align}
  \langle b'|\mathbb{E}(\widetilde{Z_2\otimes Z_1})|b'\rangle=\sum_b\langle b|Z_2\otimes Z_1|b\rangle p(b|b').
\end{align}
This yields diagonal operators for $\mathbb{E}(\widetilde{Z_2\otimes Z_1})$, $\mathbb{E}(\widetilde{Z_2\otimes\mathbbm{1}_1})$, and $\mathbb{E}(\widetilde{\mathbbm{1}_2\otimes Z_1})$, which can be expressed in terms of the noise-free operators $Z_2\otimes Z_1$, $Z_2\otimes\mathbbm{1}_1$, $\mathbbm{1}_2\otimes Z_1$, and $\mathbbm{1}$. Using the trivial equation $\mathbb{E}(\widetilde{\mathbbm{1}_2\otimes\mathbbm{1}_1})=\mathbbm{1}_2\otimes\mathbbm{1}_1$, we obtain as a final result for the noisy expectations
\begin{align}\label{eq:correlated-to-invert}
  \begin{pmatrix}
    \mathbb{E}(\widetilde{Z_2\otimes Z_1})\\
    \mathbb{E}(\widetilde{Z_2\otimes\mathbbm{1}_1})\\
    \mathbb{E}(\widetilde{\mathbbm{1}_2\otimes Z_1})\\
    \mathbb{E}(\widetilde{\mathbbm{1}_2\otimes\mathbbm{1}_1})\\
  \end{pmatrix}
  =\Omega
  \begin{pmatrix}
    Z_2\otimes Z_1\\
    Z_2\otimes\mathbbm{1}_1\\
    \mathbbm{1}_2\otimes Z_1\\
    \mathbbm{1}_2\otimes\mathbbm{1}_1\\
  \end{pmatrix}
  \quad\text{with}\quad
  \Omega_{j,k} = \sum_{b,b'}\bra b O_j\ket b\bra{b'} O_k\ket{b'}p(b|b'),
\end{align}
where $\sum_{b,b'}$ ranges over $b,b'\in\{00,01,10,11\}$, and the operators $O_j$ and $O_k$ are given by $O_1=Z_2\otimes Z_1$, $O_2=Z_2\otimes \mathbbm{1}_1$, $O_3=\mathbbm{1}_2\otimes Z_1$, and $O_4=\mathbbm{1}_2\otimes \mathbbm{1}_1$.

If the bit-flip error rate $\varepsilon=1-\min_bp(b|b)$ is below $0.05$, the matrix $\Omega$ is strictly diagonally dominant, so Eq.~\eqref{eq:correlated-to-invert} can be inverted to obtain the equivalent of Eq.~\eqref{eq:inverted} for two correlated qubits.

\subsubsection{Numerical results}

\begin{figure}[b]
    \centering
    \includegraphics[width=0.9\textwidth]{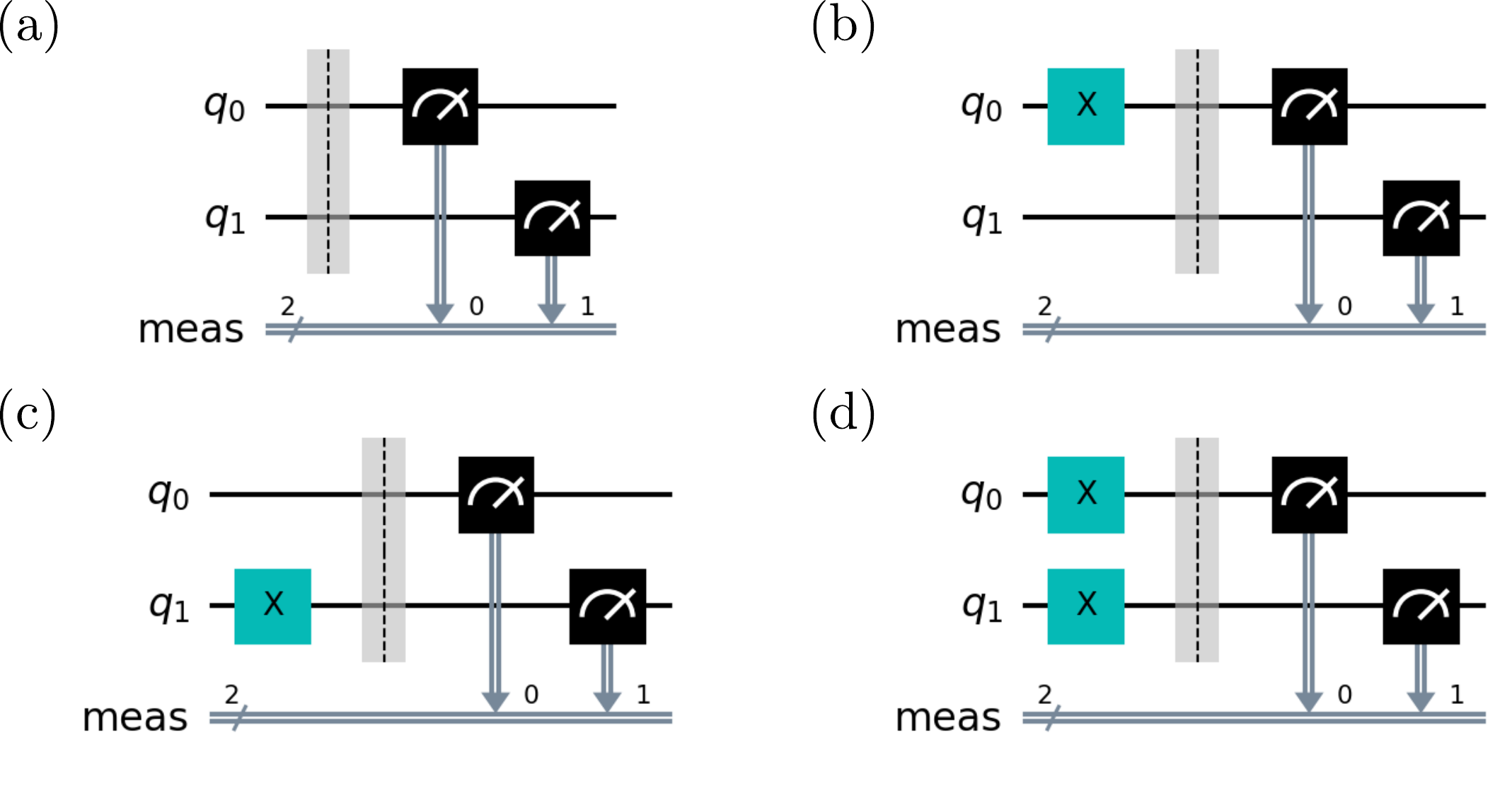}
    \caption{Illustration of the four different quantum circuits required to calibrate the bit-flip probabilities (a)~$p(b|00)$, (b) $p(b|01)$, (c) $p(b|10)$, and (d) $p(b|11)$. The green boxes denote $X$ gates, the black boxes are the final measurements, and the vertical dashed lines separate different layers of the quantum circuit.}
    \label{fig:calibration_circuits}
\end{figure}

\begin{figure}[!htp]
    \centering
    \includegraphics[width=0.7\textwidth]{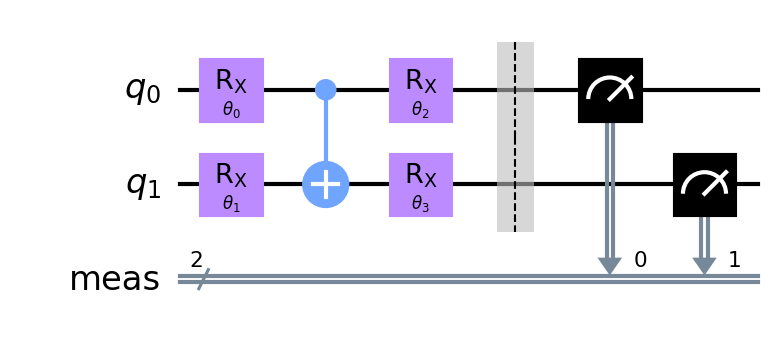}
    \caption{Parametric quantum circuit used in the experiments. The purple boxes denote single-qubit $R_X$ rotation gates with parameters $\theta_0,\dots,\theta_3$, the blue two-qubit connection is an entangling CNOT gate, the black boxes are the final measurements, and the vertical dashed lines separate different layers of the circuit.}
    \label{fig:2q_circuit}
\end{figure}

In order to test the mitigation scheme and to assess the effect of correlations, we implement our method for the case of two qubits on different quantum devices. In a first step, we calibrate the bit-flip probabilities $p(b|b')$ by repeatedly preparing each of the four possible computational basis states $\ket{b'}$ and by recording the measurement outcomes $b$ (see Fig.~\ref{fig:calibration_circuits} for details).
After calibrating the bit-flip probabilities, we run the parametric quantum circuit shown in Fig.~\ref{fig:2q_circuit} to prepare a state $\ket{\psi}$, where our choice of circuit is inspired by the typical layered structure employed in many hybrid quantum-classical algorithms, such as the variational quantum eigensolver~\cite{Peruzzo2014}. 
Subsequently, we measure the (noisy) expectation values of the four operators appearing on the left-hand side of Eq.~\eqref{eq:correlated-to-invert}. In order to obtain the (noisy) estimate of the expectation values $\mathbb{E}\bra{\psi}\widetilde{\mathds{1}_2\otimes Z_1}\ket{\psi}$, $\mathbb{E}\bra{\psi}\widetilde{Z_2\otimes \mathds{1}_1}\ket{\psi}$, and $\mathbb{E}\bra{\psi}\widetilde{Z_2\otimes Z_1}\ket{\psi}$ for a fixed set of parameters $\theta_0,\dots,\theta_3$, we have to run the circuit multiple times and collect statistics of the measurement outcomes. We refer to these number of repetitions as the number of shots $s$.  Following that, we invert Eq.~\eqref{eq:correlated-to-invert} to mitigate the effects of noise and to retrieve the true expectation values of the observables. In addition, to probe for the effect of correlated bit flips, we also use the mitigation scheme neglecting the correlations, as described in Eq.~\eqref{eq:inverted}. Comparing these results to the ones obtained with the mitigation scheme taking into account correlated bit flips allows us to assess the influence of such correlations.

To acquire statistics, we repeat the experiment described above for 1000 randomly chosen states $\ket{\psi}$, where we draw the angles $\theta_0,\dots,\theta_3$ uniformly from $[0,2\pi)$, and we monitor the mean of the absolute error
\begin{align}
    \left| \bra{\psi}\widetilde{Z_2\otimes Z_1}\ket{\psi}_\text{measured} - \bra{\psi}Z_2\otimes Z_1\ket{\psi}_\text{exact}\right|\,.
    \label{eq:error}
\end{align}
In this equation, the expression $\bra{\psi}\widetilde{Z_2\otimes Z_1}\ket{\psi}_\text{measured}$ refers to either the unmitigated measurement results for the operator $Z_2\otimes Z_1$ from the quantum device or the results obtained after applying the mitigation scheme, whereas $\bra{\psi}Z_2\otimes Z_1\ket{\psi}_\text{exact}$ refers to the exact solution obtained on a noise-free device with an infinite number of shots.

\begin{figure}[!htp]
    \centering
    \includegraphics[width=0.45\textwidth]{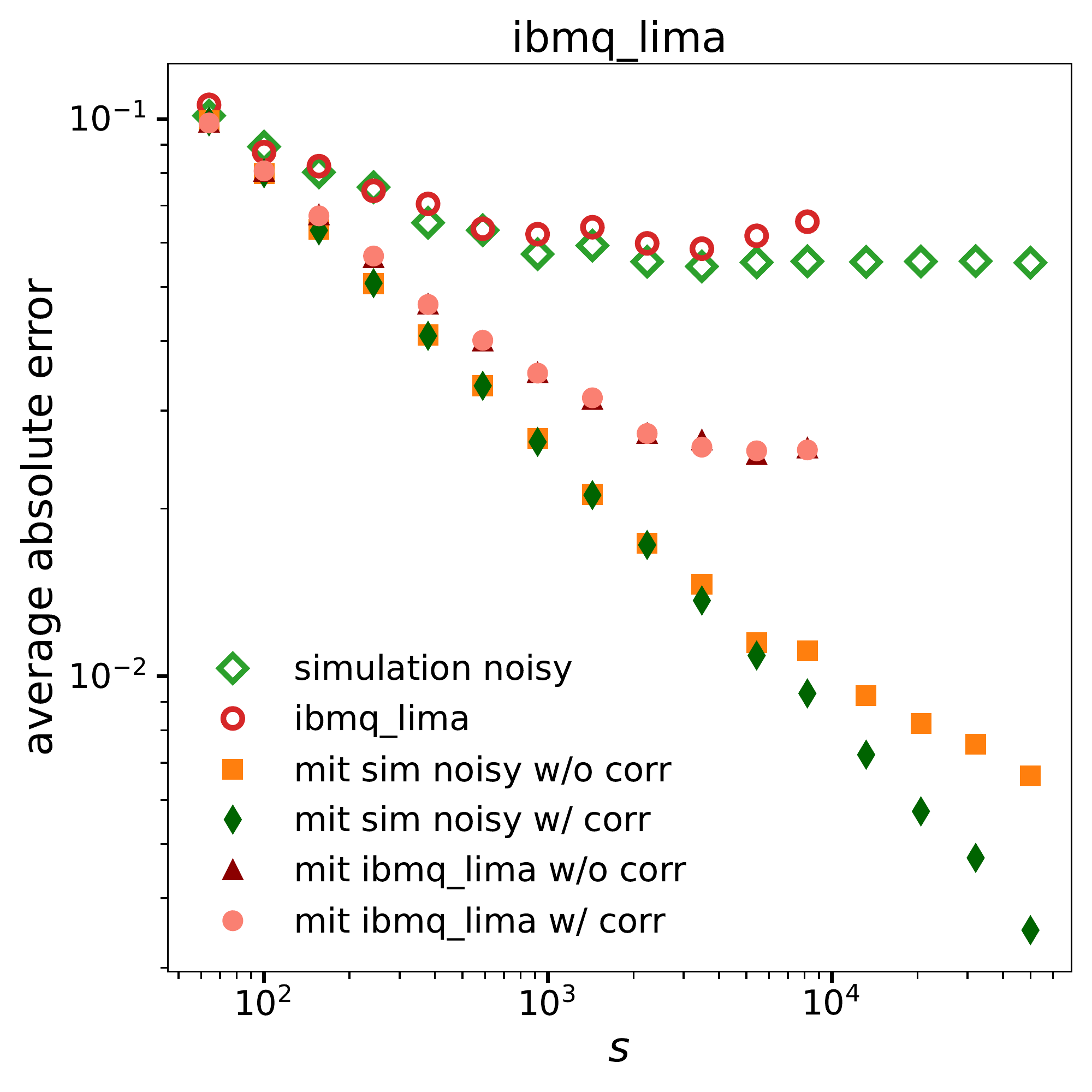}
    \put(-23,170){(a)}
    \hspace{2em}
    \includegraphics[width=0.45\textwidth]{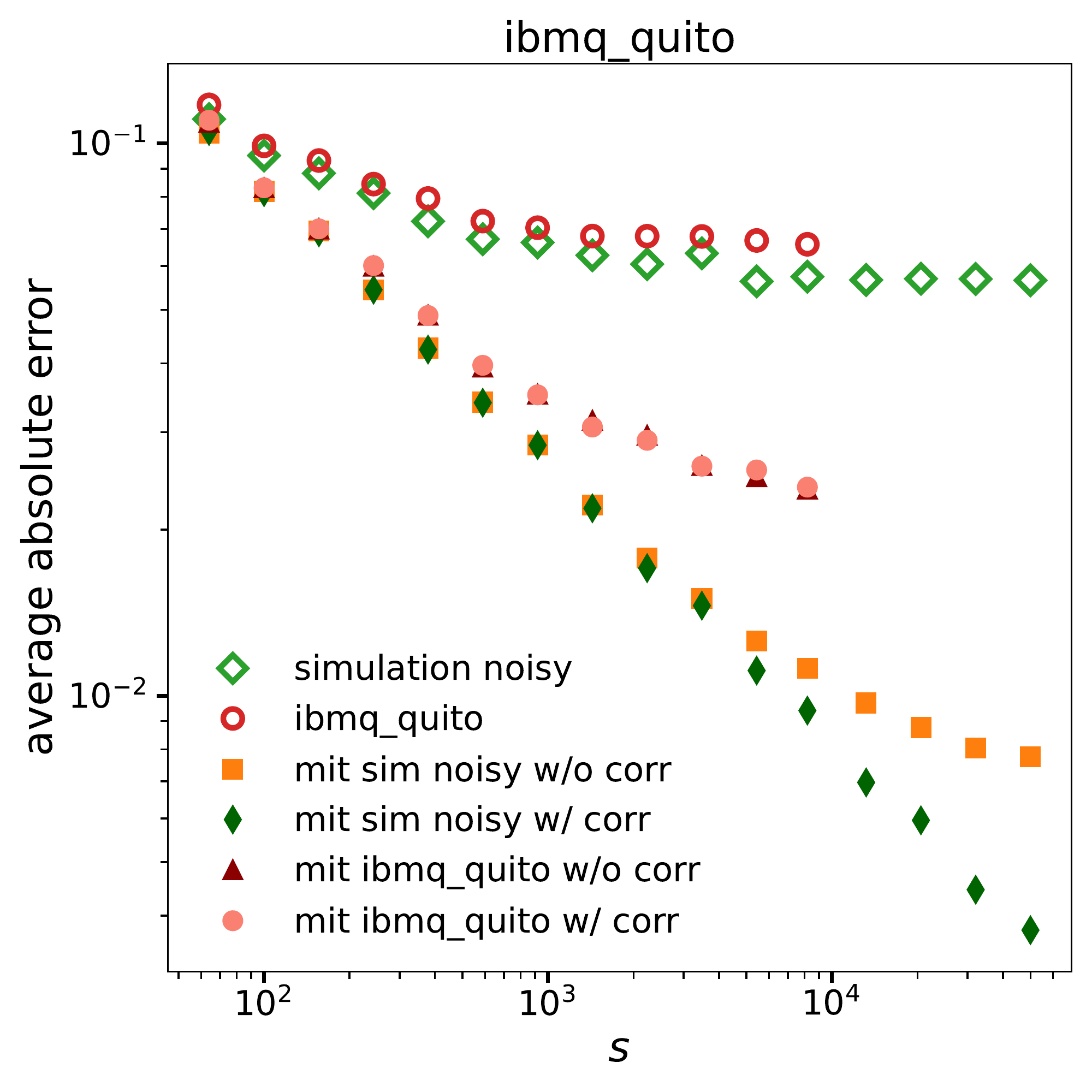}
    \put(-23,170){(b)}\\
    \vspace{1em}
    \includegraphics[width=0.45\textwidth]{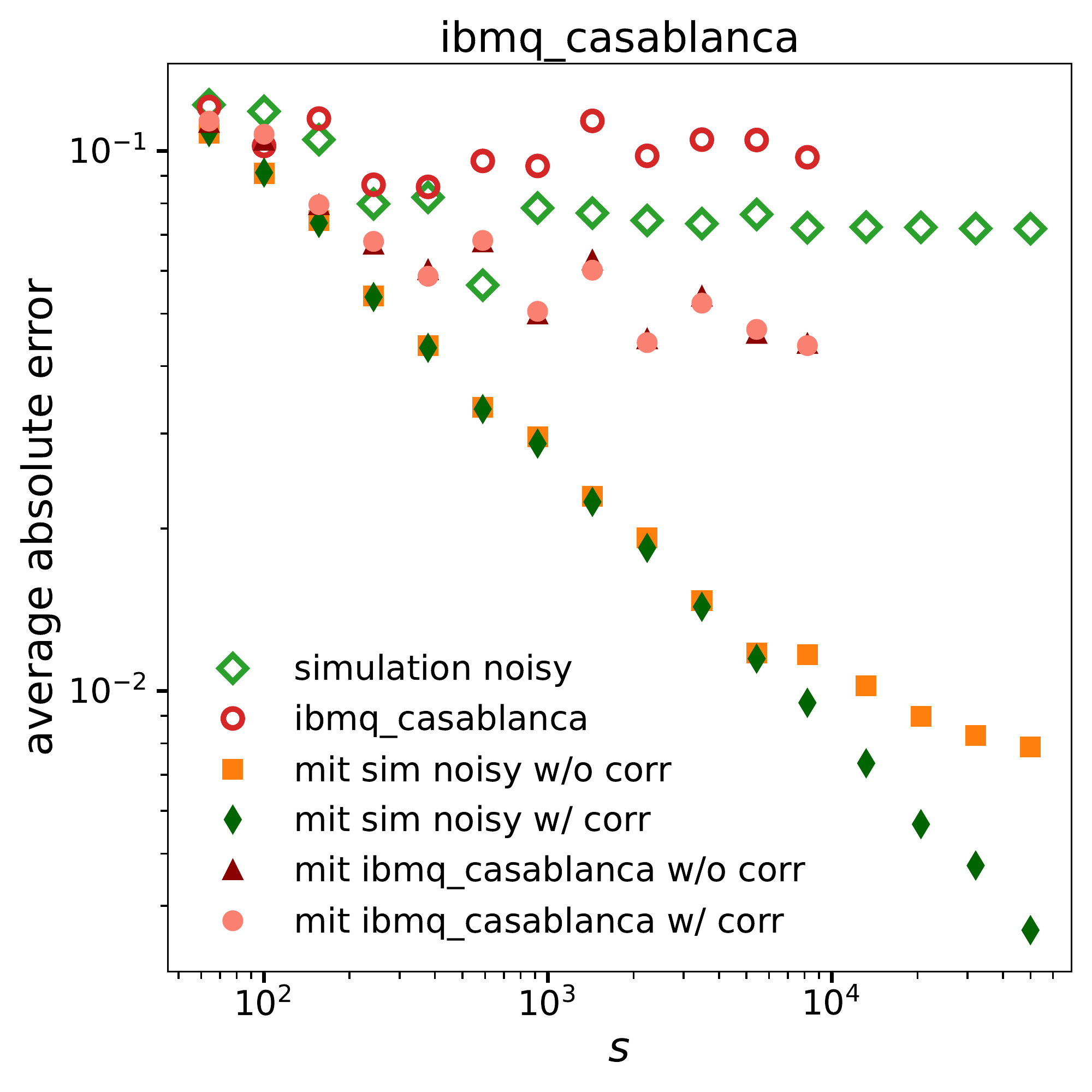}
    \put(-23,170){(c)}
    \hspace{2em}
    \includegraphics[width=0.45\textwidth]{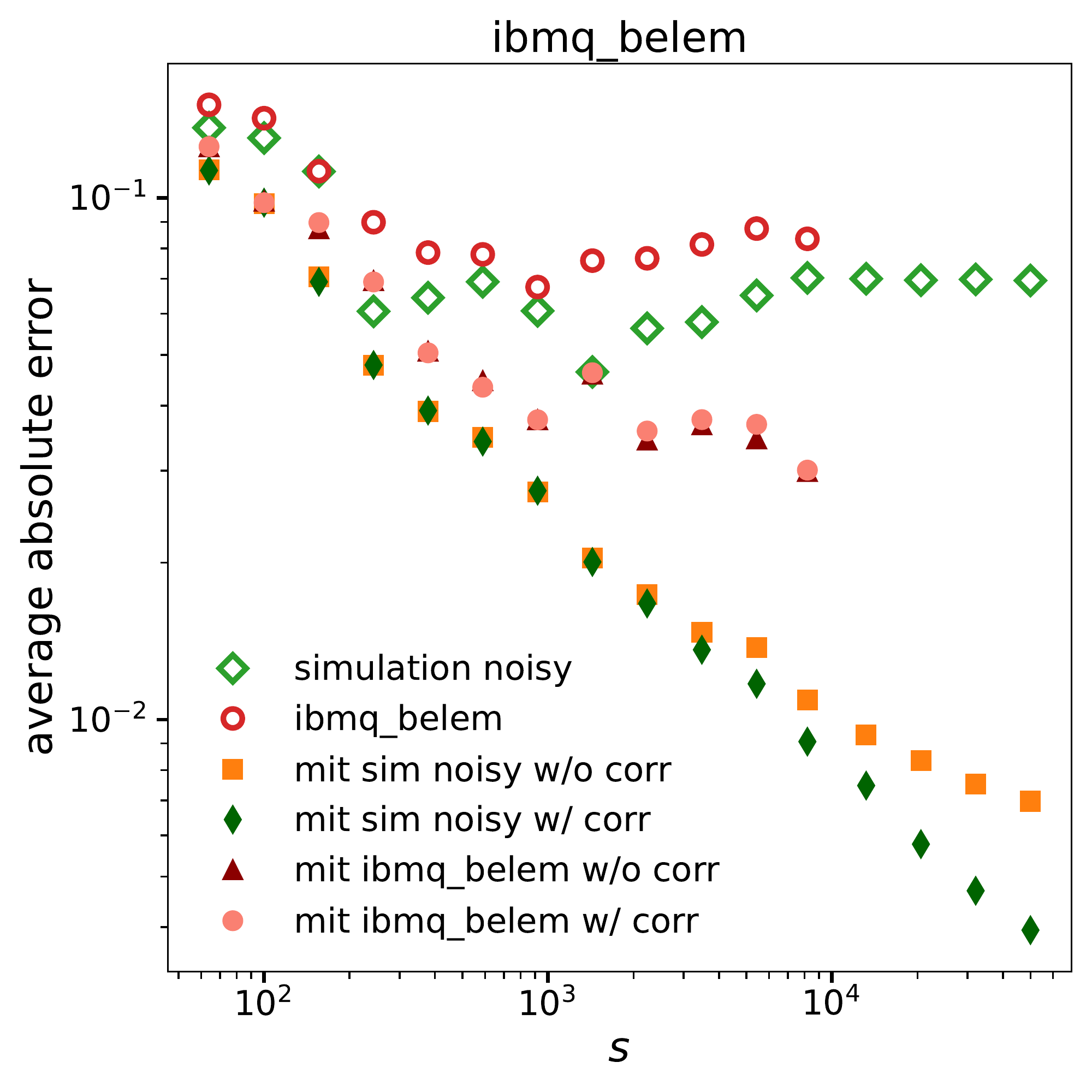}
    \put(-23,170){(d)}\\
    \caption{Average absolute errors according to Eq.~\eqref{eq:error} as a function of the number of shots $s$ for the quantum devices (a) \emph{ibmq\_lima}, (b) \emph{ibmq\_quito}, (c) \emph{ibmq\_casablanca}, and (d) \emph{ibmq\_belem}. The open symbols correspond to the unmitigated data, where the diamonds (circles) denote the data from the classical simulation (quantum hardware). The filled symbols correspond to the data obtained after applying the mitigation procedure (abbreviated ``mit'' in the legend). The orange squares (dark green diamonds) represent the mitigated noisy simulation data without (with) taking two-qubit correlations into account. The dark red triangles (rose circles) correspond to the mitigated hardware data without (with) taking two-qubit correlations into account. The maximum number of shots on hardware devices is limited to 8192; thus, only simulation data is available for values of $s$ exceeding this number.}
    \label{fig:results}
\end{figure}

In Fig.~\ref{fig:results}, we show our results for the average error as a function of the number of shots obtained on the different quantum devices \emph{ibmq\_lima}, \emph{ibmq\_quito}, \emph{ibmq\_casablanca}, and \emph{ibmq\_belem}, which are all IBM Quantum Falcon Processors. In addition to the hardware data, we also provide results from a classical simulation mimicking a quantum device with readout errors only for reference. The bit-flip probabilities for the classical simulation are set to the values obtained from the calibration on quantum hardware.
Focusing on the classical simulation first, we observe that the absolute average error for the unmitigated data initially decreases slightly with increasing number of shots $s$, before quickly reaching a plateau around $7\times10^{-2}$. In particular, the plateau occurs at a number of shots much smaller than the one that can be executed on real hardware, thus demonstrating that readout errors severely limit the accuracy that can be obtained. Applying the mitigation procedure that neglects the correlations (see Eq.~\eqref{eq:inverted}) significantly improves the results, as the orange squares in Fig.~\ref{fig:results}(a)-(d) reveal. However, around a value of $s=10^4$ we observe that the absolute value of the average error again tends to stagnate, hence indicating that we have reached the level of accuracy at which the correlated bit flips between qubits become significant and thus the results do not further improve. In contrast, when using the mitigation scheme from Eq.~\eqref{eq:correlated-to-invert} that takes into account the correlations, there is no trend towards a plateau of the average absolute error, and our data show a polynomial decay $\propto1/\sqrt{s}$ throughout the entire range of shots we study. This scaling of the average error with the number of shots is expected for the ideal noise free case~\cite{Funcke_2021}, thus showing that the mitigation scheme allows for recovering the noise-free case.

Turning to the data obtained on \emph{ibmq\_lima} and \emph{ibmq\_quito} without error mitigation (see Figs.~\ref{fig:results}(a) and \ref{fig:results}(b)), we observe good agreement with the simulation taking into account readout errors only, hence indicating that readout errors have a quite significant contribution to the overall error on these devices. Applying the error mitigation again yields a substantial improvement for the results. After an initial decrease with the number of shots, the average errors of the mitigated data show a trend towards a plateau around $3\times 10^{-2}$. Since the inherent statistical fluctuations of the projective measurements should decrease with an increasing value of $s$, and our mitigation scheme allows for dealing with readout errors, this hints towards other noise sources becoming dominant at this stage, eventually limiting the accuracy that can be reached on the quantum device. Interestingly, the results obtained from the readout error mitigation scheme that takes correlated bit flips into account are essentially identical to those from the method that neglects the correlations. Thus, our results suggest that for the level of accuracy that can be reached on real quantum devices, correlations between the readout errors of qubits do not play a significant role.

Looking at the hardware results in Figs.~\ref{fig:results}(c) and \ref{fig:results}(d) from our simulations on \emph{ibmq\_casablanca} and \emph{ibmq\_belem}, the picture is qualitatively similar. However, for these devices we observe larger deviations between the classical simulation that takes only readout errors into account and the actual hardware data. This is giving an indication that other sources of noise have a more significant contribution on these devices compared to \emph{ibmq\_lima} and \emph{ibmq\_quito}. Using the readout error mitigation again improves the results, albeit the effect for these devices is less drastic, due to noise other than readout errors. Moreover, also for this case we observe that both mitigation schemes yield essentially identical results, as a comparison between the dark red triangles and the rose dots shows. Consequently, correlated readout errors do not play a significant role and can be neglected.

\section{Conclusion}

In Refs.~\cite{Funcke:2020olv,Funcke:2021aps}, we have introduced a readout error 
mitigation scheme which is efficient, scales only polynomially in the number of qubits, and can be 
practically implemented. In the cited papers, we have also performed 
experiments on IBMQ hardware and demonstrated the feasibility of our method. 
Further experiments, including the variances of the error and a comparison 
between quantum hardware of Rigetti and IBMQ have been performed 
in Ref.~\cite{Alexandrou:2021ynh} of this conference.  

Although the general case that correlations between qubits in the readout process can occur 
has been discussed in Ref.~\cite{Funcke:2020olv}, no numerical experiments had been 
performed there. In these proceedings, we have filled this gap and have successfully
applied our readout error mitigation scheme taking correlations into account. 
The numerical experiments have been performed on several IBMQ hardware devices 
for two qubits. 
When comparing results with and without adding correlations, we obtained 
similar outcomes, suggesting that correlations can be neglected. 
This justifies the assumption in Refs.~\cite{Funcke:2020olv,Funcke:2021aps} to consider 
only cases without correlations.

\acknowledgments
We thank Giovanni Ianelli and Tom Weber for many useful discussions.
Research at Perimeter Institute is supported in part by the Government of Canada through the Department of Innovation, Science and Industry Canada and by the Province of Ontario through the Ministry of Colleges and Universities. 
L.F.\ is partially supported by the U.S.\ Department of Energy, Office
of Science, National Quantum Information Science Research Centers,
Co-design Center for Quantum Advantage (C$^2$QA) under contract number
DE-SC0012704, by the DOE QuantiSED Consortium under subcontract number
675352, by the National Science Foundation under Cooperative Agreement
PHY-2019786 (The NSF AI Institute for Artificial Intelligence and
Fundamental Interactions, http://iaifi.org/), and by the U.S.\
Department of Energy, Office of Science, Office of Nuclear Physics under grant contract numbers DE-SC0011090 and DE-SC0021006.
S.K.\ acknowledges financial support from the Cyprus Research and Innovation Foundation under project ``Future-proofing Scientific Applications for the Supercomputers of Tomorrow (FAST)'', contract no.\ COMPLEMENTARY/0916/0048. 
G.P.\ is supported  by project NextQCD, co-funded by the European Regional Development Fund and the Republic of Cyprus through the Research and Innovation Foundation (EXCELLENCE/0918/0129) and POST-DOC/0718/0100. Partial support is provided by the Marie Skłodowska-Curie European Joint Doctorate program STIMULATE of the European Commission under grant agreement No 765048. We acknowledge the use of IBM Quantum services for this work. The views expressed are those of the authors, and do not reflect the official policy or position of IBM or the IBM Quantum team.

\bibliographystyle{JHEP}
\bibliography{Papers}

\end{document}